\documentclass[a4paper,10pt,pra,aps,twocolumn,showpacs]{revtex4}
\usepackage{amsmath,amssymb,graphicx}
\usepackage{datetime,hyperref}

\newcommand{\figref}[1]{Fig.~\ref{#1}}

\newcommand{\micron}{\ensuremath{\mu{\rm m}}}

\begin{document}

\title{{Experimental Evidence for Inhomogeneous-Pumping and Energy-Dependent Effects in Photon Bose-Einstein Condensation}}

\author{J. Marelic}
\author{R. A. Nyman}\email[Correspondence should be addressed to
]{r.nyman@imperial.ac.uk} \affiliation{Centre for Cold Matter,
Blackett Laboratory, Imperial College London, Prince Consort Road, SW7
2BW, United Kingdom}

\date{\today}
\begin{abstract}
	Light thermalised at room temperature in an optically pumped, dye-filled microcavity resembles a model system of non-interacting Bose-Einstein condensation in the presence of dissipation. We have experimentally investigated some of the steady-state properties of this unusual state of light and found features which do not match the available theoretical descriptions. We have seen that the critical pump power for condensation depends on the pump beam geometry, being lower for smaller pump beams. {Far below threshold, both intracavity photon number and thermalised photon cloud size} depend on pump beam size, with optimal coupling when pump beam matches the thermalised cloud size. We also note that the critical pump power for condensation depends on the cavity cutoff wavelength and longitudinal mode number, which suggests that {energy-dependent thermalisation and loss mechanisms are important}.
\end{abstract}


\maketitle

{The decision to categorise an experimentally observed phenomenon as Bose-Einstein condensation (BEC) goes hand-in-hand with the consensus microscopic description. }For example, the popular definition of BEC by Penrose and Onsager~\cite{Penrose56} of extensive or macroscopic occupancy by identical bosons of a single quantum state was chosen to extend the original idea of Bose and Einstein to interacting particles, implicitly assuming homogeneity, in their case superfluid Helium.

In general, BEC at thermal equilibrium arises {because the Bose-Einstein distribution} diverges when the chemical potential is at least equal to the energy of ground state. In dissipative, non-equilibrium condensation of {exciton-polaritons} in semiconductors~(e.g. \cite{Kasprzak06, Deng06, Balili07}) or of polaritons in organic molecules~\cite{Daskalakis14, Plumhof14}, the system may be effectively homogeneous, so the Penrose and Onsager definition of BEC applies, but thermal equilibrium is not always strongly established. {In these cases, BEC is widely accepted when thermal equilibrium is experimentally demonstrated to be a good description, and a macroscopic population is observed in the lowest energy state, despite the strong interactions.}

{Photons thermalised in a dye-filled microcavity probably have the weakest interactions of any system to have exhibited BEC. In this intrinsically inhomogeneous system, thermal equilibrium and macroscopic occupancy of the ground state are the usual criteria for BEC, and both have been observed despite the dissipation~\cite{Klaers10a, Klaers10b}, so BEC is uncontroversially assigned. Interactions are so weak, that questions have been asked about the mechanism by which the condensate forms~\cite{Snoke13}. There has been considerable recent activity developing microscopic models of this physical system, but most of the models, e.g. by Kruchkov~\cite{Kruchkov14}, assume that near-thermal-equilibrium conditions hold.}

{
Using principals of detailed balance~\cite{Klaers12} and hierarchical maximum entrance~\cite{Sobyanin12, Sobyanin13}, fluctuations of the condensate population about the thermal equilibrium have been predicted and subsequently observed~\cite{Schmitt14}. Likewise, low-energy excitations about the condensate mean-field such as the Bogoliubov dispersion~\cite{Zhang12, Nyman14} have also been calculated. Phase fluctuations can only be predicted by fully-quantised models including dissipation~\cite{Leeuw13, Chiocchetta14}.}

One published model~\cite{Kirton13, Kirton14private} looks at the limits of the thermalisation process itself, and hence can state when BEC is and is not a good description. When thermalisation is slower than loss, threshold may still be reached, but the macroscopically occupied mode may no longer be the ground state. In other words, BEC breaks down when thermalisation breaks down. Validation of this and all the other models requires new experimental evidence. 

We report here our own observations of dye-filled microcavity photon condensation in the steady state. We have seen that the critical pump power varies strongly with the pump beam geometry, in stark contrast to the predictions of a simple, equilibrium model~\cite{Klaers10b, Kruchkov14}. We demonstrate that even below threshold the model is incorrect: the thermalised cloud size and photon number are also pump-geometry dependent. We also measure critical pump power as ground-state energy and overall cavity length vary, and we explain our observations through energy-dependent losses. These steady-state features should be explained described by any successful model of photon BEC.

We note that very recent experiments have looked at aspects of the time-resolved behaviour of photon thermalisation~\cite{Schmitt15}, and the crossover to lasing when thermalisation fails.

\begin{figure} 
	\includegraphics[width=0.48\textwidth]{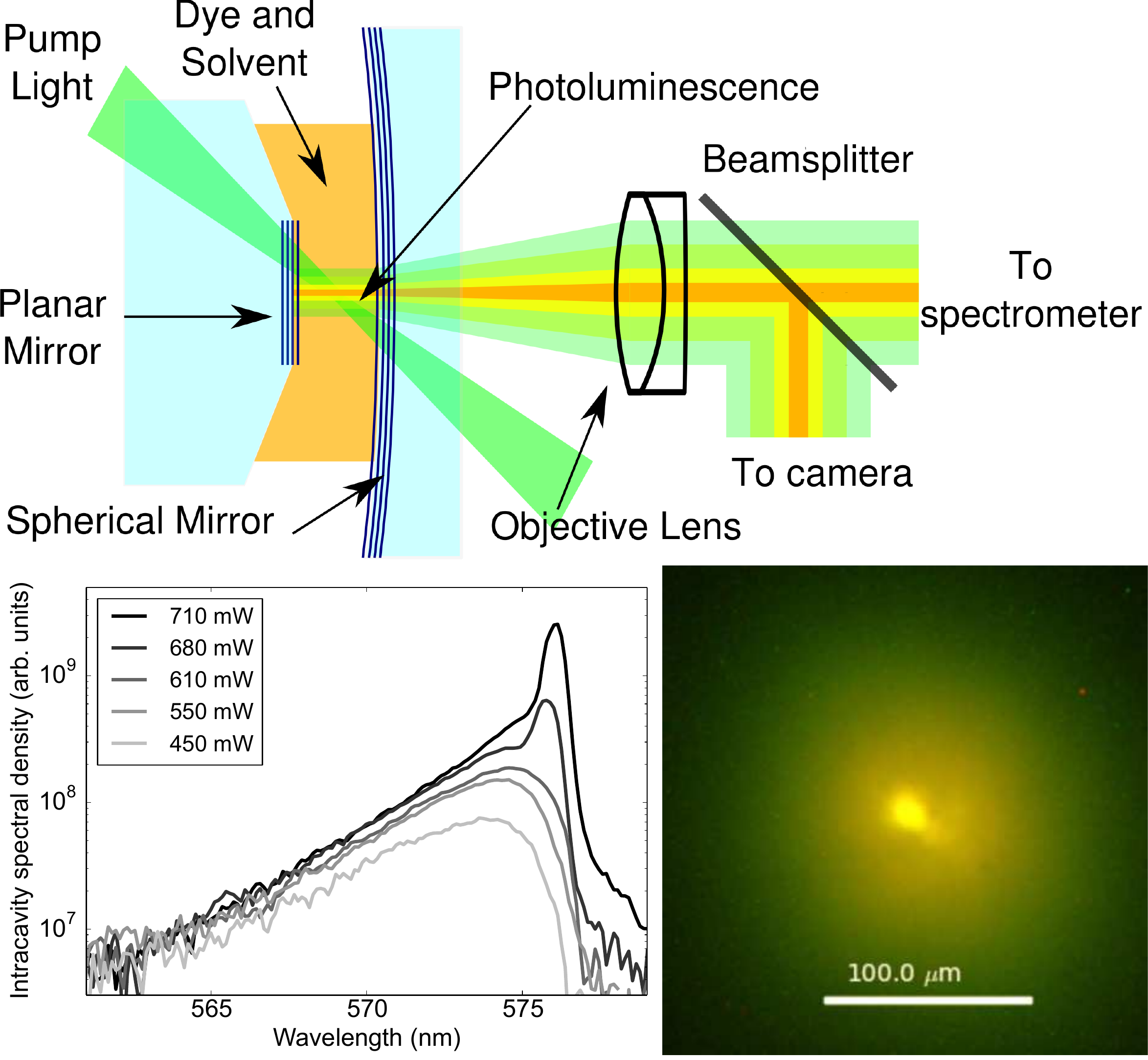}
	\caption{Schematic diagram of the apparatus (top) and some example data demonstrating BEC of photons in a dye-filled microcavity. Bottom left: spectra at varying intracavity pump powers, showing the saturation of the excited-state population and then condensation into the lowest-energy mode available. Bottom right: an image of a condensate just above threshold, in real colours, albeit with the intensity adjusted for visibility when printed. }
	\label{fig:making bec}
\end{figure}

{Our experimental method is similar to that of }Klaers \textit{et al}~\cite{Klaers10a,Klaers10b}. A fluorescent dye, Rhodamine 6G dissolved in Ethylene Glycol at 2~mM concentration, is held by surface tension between two dielectric mirrors placed \mbox{1--2~\micron} apart, as shown in \figref{fig:making bec}(top). One of the two mirrors is spherical with a radius of curvature $R=0.25$~m and the other is planar, cut down to 1~mm diameter. We pump the dye {incoherently using $\lambda_{pump}=532$~nm light}, passing through the dielectric mirror at a transmission maximum angle around 37$^\circ$ to the normal. To prevent shelving of the dye in the triplet state, we pulse the pump on for 500~ns at a repetition rate of 500~Hz.

We collect the light leaking through the cavity mirrors, the photoluminescence, using a 50~mm focal length, 25~mm diameter, achromatic doublet as an objective lens in an afocal setup: an image is effectively formed at infinity. The light is split and imaged onto a camera and a commercial spectrometer, whose entrance slit is easily replaceable. The length of the cavity is controlled using a piezoelectric actuator, and stabilised by {reference to a collimated helium-neon laser} (at 633~nm) as observed by a secondary camera. Pump and stabilisation wavelengths are separated from photoluminescence using a combination of dichroic, notch and short-pass filters.

We show an example image of BEC of photons in \figref{fig:making bec}(bottom right). 
The colours correspond nearly to those observed by the camera. 
The thermal component is the broad Gaussian cloud around the condensate, which shows up as a bright central spot. In the spectrum, \figref{fig:making bec}(bottom left), the condensation shows up as an increase in the population of the lowest energy state, at the cavity cutoff. The thermal component is compatible with a room-temperature thermal distribution, although here the 50~\micron\ spectrometer entrance slit cuts off some of the higher energy components.

We determine the threshold power using the {deviation of the spatial variation of photoluminescence from a Gaussian near the centre}. This measure has proven to be precise and robust and is as well performed by eye as by any quantitative measure we have tried, e.g. output power as a function of pump power, or fitting of the spectra.


The simplest theory of thermalised BEC of dye-filled microcavity photons{, as used in Ref.~\cite{Klaers10b},} assumes that a number of photons are trapped in the cavity at thermal equilibrium, and that above the critical number a condensate will form. The critical particle number per spin state for equilibrium Bose-Einstein condensation in a symmetric two-dimensional (2D) harmonic oscillator is $N_C = g \frac{\pi^2}{6}\left( \frac{k_B T}{\hbar \Omega}\right)^2$ where $T$ is the temperature, $\Omega = (c/n_L)\sqrt{1/LR}$ is the angular trapping frequency for {photons a cavity of length $L$, filled with a medium of refractive index} $n_L$, and $g$ is the spin degeneracy.

The photon number stored in a dye-filled microcavity is equal to the light power absorbed from the pump, $P_{abs} = P_{pump} n_{mol} \sigma_{abs} \frac{\lambda_0}{2n_L}(q-q_0)$ {(assuming that the pump couples well to the cavity)}, times the time the light circulates for, {$\tau_{cav} = \frac{F\lambda_0}{c}q$}, divided by the {typical energy per photon absorbed, $h c / \lambda_{pump}$}. Here, $P_{pump}$ is the pump light power inside the cavity, $n_{mol}$ is the dye-molecule volumetric number density, $\sigma_{abs}$ is the absorption cross-section for light at the pump wavelength. The lowest-energy mode of the cavity at the cutoff wavelength $\lambda_0$ is in the $q^{\rm th}$ longitudinal mode, the parameter $q_0$ indicates how far the light penetrates the surface of the mirrors, and $F$ is the typical number of round trips light will make in the cavity before decay {(i.e. equal to the finesse divided by $\pi$)}. {The intracavity photon number is then 
\mbox{$
	N_{cav} = {P_{pump} n_{mol} \sigma_{abs}  \frac{F\lambda_0^2 \lambda_{pump}}{2n_L h c^2}}\times q(q-q_0)
	\label{eqn:Ncav}
$.}}
In such a cavity, a thermal cloud, below threshold, is expected to have a root-mean-square size of \mbox{$\sigma_{th}=\sqrt{{k_B T q \lambda_0^2 R}\,/\,{2 h c n_L}}$}. 

The critical pump power is then predicted to be:
{
\mbox{$
	P_{C} 
		= g\frac{2 \pi^4 R  (k_B T n_L)^2}{3 \, h \,n_{mol} \,\sigma_{abs} F \lambda_{pump}}\times \frac{1}{\lambda_0\,(q-q_0)}
$.}}
For a given longitudinal mode, the critical pump power is expected to decrease with increasing cutoff wavelength, albeit only slightly. The critical power also decreases as the longitudinal mode number is increased. {We note that the predictions of this model are independent of the spatial distribution of pump light. It is assumed that all the pump light absorbed is fully thermalised, and a single lifetime is assigned to all cavity modes at all energies.}


The pump light is focussed into the cavity through planar mirror. To measure the pump spot size, we lengthen the cavity to around 30~\micron, and observe only the photoluminescence at 625~nm and more, to eliminate the possibility of observing thermalised light. We fit a 2D Gaussian and infer two width parameters (root-mean-square size, which is half the beam waist parameter sometimes used in optics), since our spots are often elliptical or even more distorted. We give uncertainties in spot size which are half the difference between the two width parameters.


\begin{figure}
	\includegraphics[width=0.35\textwidth]{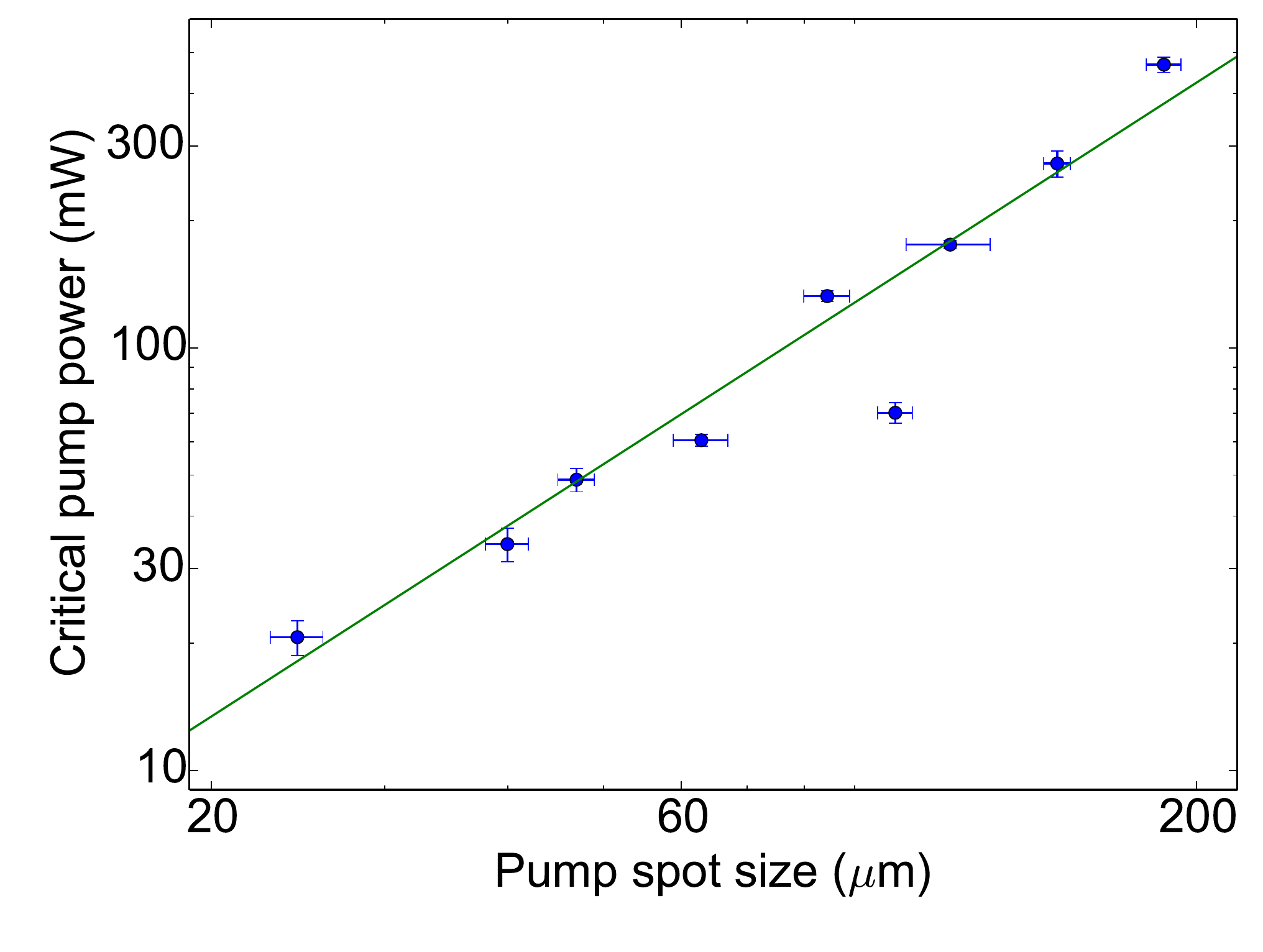}
	\caption{Critical intracavity pump power variation with pump spot size for the 8\textsuperscript{th} longitudinal mode and cavity cutoff wavelength 586~nm. The line is a guide to the eye, proportional to \mbox{(spot size)$^{1.5}$}. Error bars are from the differences in inferred size of the two principal axes of a two-dimensional Gaussian fitted to the pump spot images.}
	\label{fig:Pcrit vs spot size}
\end{figure}

The theory we have presented assumes that how photons are pumped into the cavity is irrelevant, so the critical pump power should be independent of pump geometry. In \figref{fig:Pcrit vs spot size} we present experimental observations of critical pump power as a function of pump spot size which directly contradict the prediction {of our simple model, which gives around 1~mW for these parameters. Clearly the pump is not well coupled to the cavity photons.} The data can be fitted well to a power law, $P_C \propto ({\rm spot\, size})^{1.5 \pm 0.1}$, although there is no reason to believe that the power law extends beyond the range of our measurements. The critical pump intensity decreases with increasing pump spot size.

There are two main saturation mechanisms in dye fluorescence, stimulated emission and pumping into a dark state, neither of which can explain our observations. The stimulated-emission saturation intensity for absorption at our pump wavelength is approximately $I_{sat} = h c /{\lambda \sigma_{abs} \tau_{sp}} = 4\times10^9$~W~m$^{-2}$, taking {the spontaneous emission lifetime as} \mbox{$\tau_{sp}=4$~ns}~\cite{Magde99,Schaefer}. The highest intensity we find at threshold is {$6\times 10^6$~W~m$^{-2}$}; stimulated emission is a negligible effect. The rate of non-radiative events, mostly intersystem crossing into the triplet state, is found from the fluorescence quantum yield, $\Phi$, about 95\% for rhodamine-6G in polar organic solvents\cite{Magde02}. The typical timescale for these events to occur with our weak pump intensities is {$\tau_{ST}=\tau_{sp}\frac{I_{sat}}{I(1-\Phi)} > 50~\mu$s}, much longer than our pulses, so singlet depopulation during a single pulse is negligible. Recovery from the triplet state is known to be no slower than $400~\mu$s~\cite{Zondervan03}, implying that all molecules return to the singlet state in the between pulses.


{Since saturation mechanisms are not responsible, we suppose that spatial redistribution of photons from the pump to the thermalised distribution is not fully efficient. It is worth noting that our smallest pump spots, 15~\micron, are not much larger than the smallest cavity mode, whose typical size is 6~\micron.} Where threshold behaviour is seen, independent of the pump geometry, it is always the lowest transverse mode which is macroscopically occupied. Along with the thermal excitations seen in both the images and the spectra, this feature points to BEC being a good description of the system\cite{Klaers10b, Kirton13}.


Having established that pump spot size affects the critical power, we now look for indications that below threshold the thermalisation behaviour also shows pump geometry dependence. We have observed both the number of thermalised photons and their spatial distribution as pump spot size varies, for pump power well below threshold. 

We infer the intracavity photon number $N_{cav}$ by measuring the output light power and accounting for the mirror transmission and cavity round-trip time:
\mbox{$
	N_{cav} = P_{measured} \times \frac{q \lambda^2}{h c^2 T_M(\lambda)}
$}.
\begin{figure} 
	\includegraphics[width=0.35\textwidth]{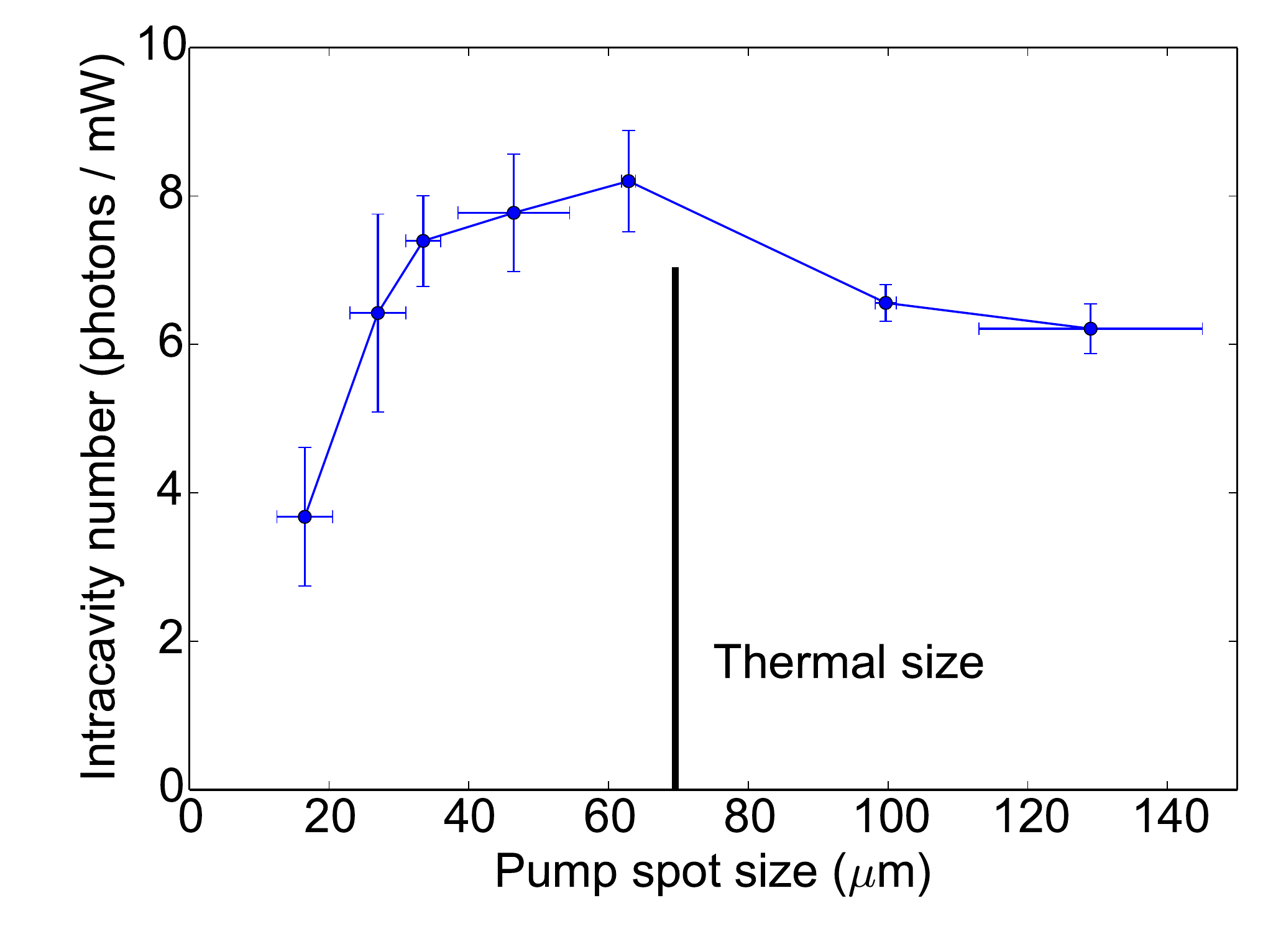}
	\caption{Cavity photon number normalised to intracavity CW pump power well below threshold, as pump spot size varies. The thermalised cloud is observed in the 8\textsuperscript{th} longitudinal mode with {cavity cutoff} 590~nm. Horizontal error bars are from the differences in inferred size of the two principal axes of a two-dimensional Gaussian fitted to the images. Vertical error bars from variation among three series of measurements spanning a factor 15 in pump power.}
	\label{fig: Ncav vs spot size}
\end{figure}
The transmission $T_M(\lambda)$ is a slowly-varying function of the wavelength. We have calibrated the normal-incidence transmission of our mirrors at two wavelengths (532 and 568~nm), and gathered wavelength variation information using fluorescence spectroscopy for each of our mirrors. When compared to the reflection specification given by the manufacturer (Ultrafast Innovations), there is a shift in the wavelengths, probably due to an unwanted tilt during coating, and an overall scaling.

There are no predictions for how the inhomogeneous pump beam couples to thermalised intracavity photons. Our observations in \figref{fig: Ncav vs spot size} indicate variations in coupling efficiency up to a factor about 2{, although always much lower efficiency than our simple model, which predicts about 20\,000 photons per mW of pump. One explanation for this poor pumping efficiency may be the large angle of incidence of the pump, giving pump photons far greater in-plane momenta than thermally accessible.} The observations are for continuous-wave (CW) pumping, always less than 20~\% of critical power. We have performed experiments over a span of a factor 15 in pump power, and observed no systematic power-dependent effects, ruling out saturation phenomena. The standard deviations over the multiple experiments are incorporated into the data shown in \figref{fig: Ncav vs spot size} as standard error bars. The optimal coupling between pump and thermalised light is found when the pump spot approximately matches the expected thermal spatial distribution of photons. {It seems that the pump couples to spatially matched modes, and that re-distribution is not fully efficient, supporting the explanation we have for the threshold behaviour, as in \figref{fig:Pcrit vs spot size}. Larger pump spots couple more weakly, and it is possible that some of the pump light strikes damaged regions of our mirrors. Smaller pump spot sizes couple very much more weakly.} Combined with the results of \figref{fig:Pcrit vs spot size}, we conclude that the threshold photon number decreases rapidly for decreasing pump spot size below the typical thermal cloud size, faster than linearly.


\begin{figure} 
	\includegraphics[width=0.35\textwidth]{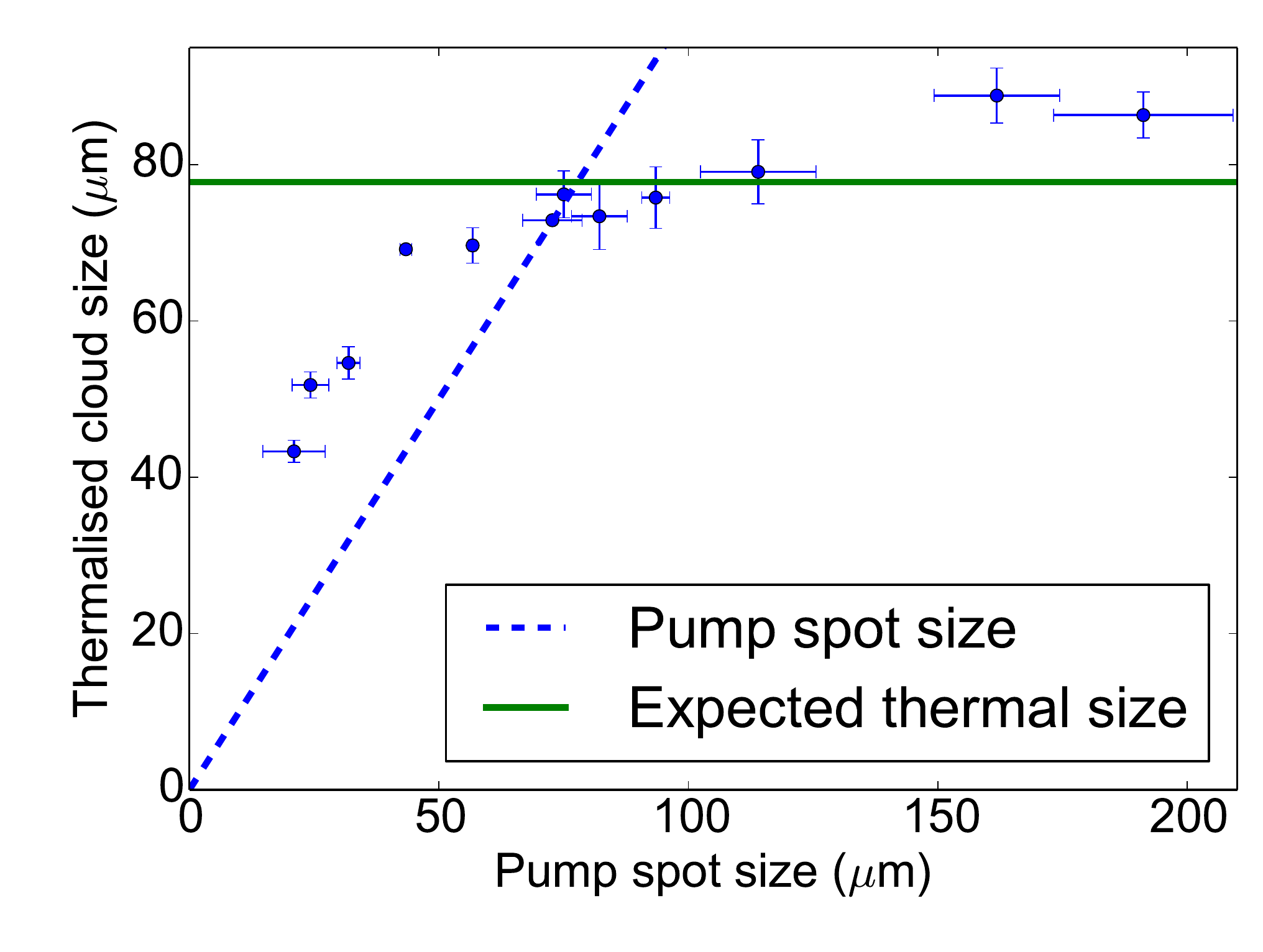}
	\caption{Thermalised photon cloud size variation with pump spot size for CW pump well below threshold. For comparison, the expected thermal cloud size for room temperature is plotted, alongside a line equal to the pump spot size. Error bars are from the differences in inferred size of the two principal axes of a two-dimensional Gaussian fitted to the images. The thermalised cloud is observed in the 10\textsuperscript{th} longitudinal mode with {cavity cutoff }590~nm.}
	\label{fig:thermal spot size}
\end{figure}

A second surrogate measure of thermalisation, cloud size, gives further information about the coupling between pump and thermalised photons. We measure the thermalised cloud size by fitting a 2D Gaussian to the photoluminescence from the cavity. Uncertainties in the data are from differences between the characteristic sizes fitted in the two dimensions. In \figref{fig:thermal spot size} we see that for large pump spots the cloud is always somewhere between the pump spot size and the expected thermal size, with the thermalisation being more dominant for larger pump spots. The implication is that {light from }larger pump spots thermalises better than small spots, but that pumping with a spot the same size as the thermal cloud gives optimised coupling.


Thermalisation depends on the scattering rate being faster than the cavity loss rate, both of which are wavelength dependent. Photon scattering from the dye decreases exponentially with increasing wavelength, and our cavity mirrors have maximum reflectivity at about 550~nm. Thermalisation is then less effective at longer wavelengths. We have made below-threshold, CW, cloud-size measurements for {cavity cutoff wavelengths} from 575--610~nm at concentrations from 0.02--2 ~mM. Under those circumstances, the rate of scattering, hence the thermalisation rate, varies by a factor of 20\,000. For small pump spots we see no systematic variation in cloud size with scattering rate. For large pump spots there is some evidence that higher scattering rates (high concentration, short cutoff wavelength) are associated with cloud sizes that better match the expected thermal cloud size.

\begin{figure} 
	\includegraphics[width=0.35\textwidth]{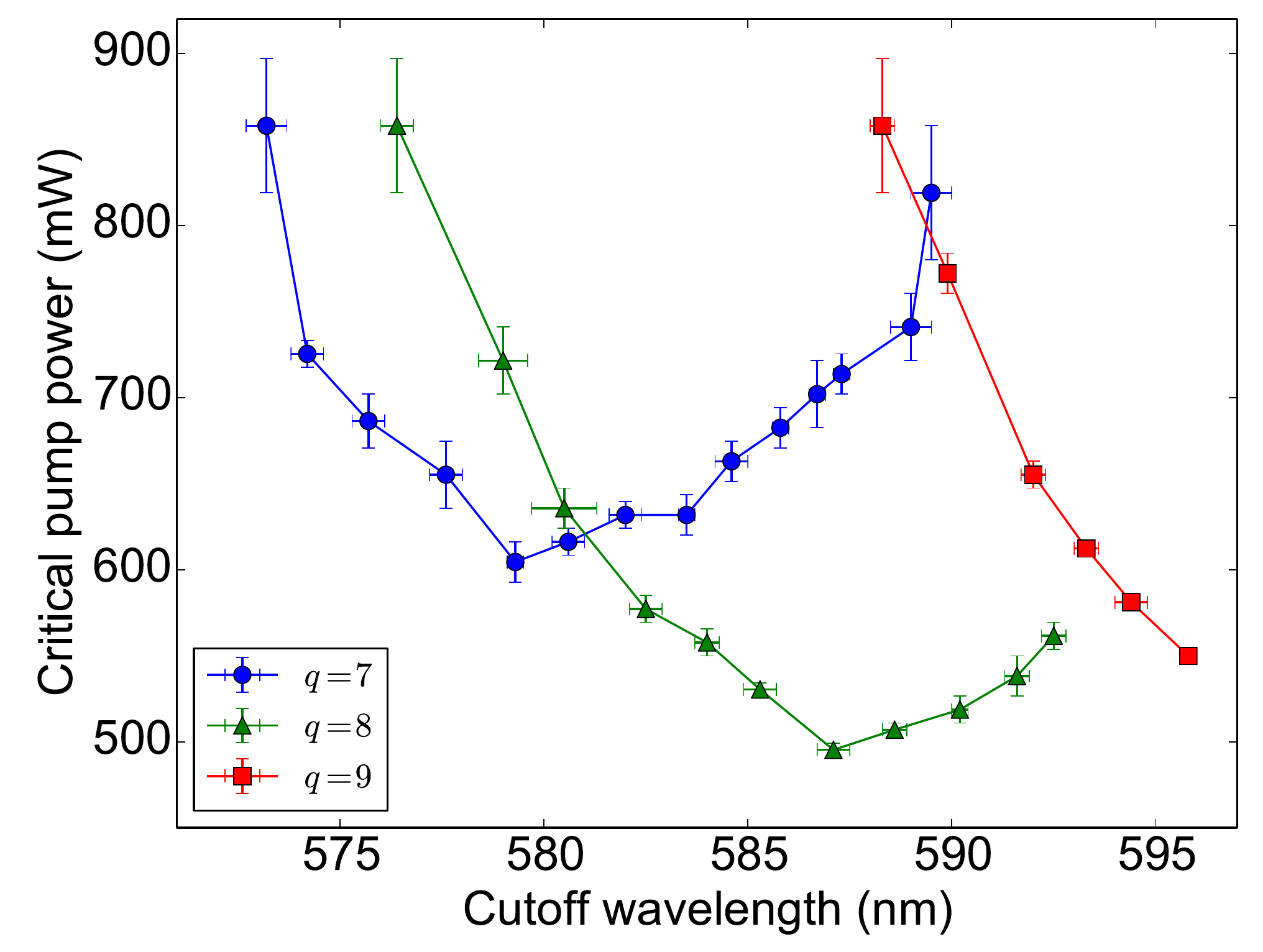}
	\caption{Critical power variation with cavity cutoff wavelength (the wavelength at which BEC appears), for various longitudinal mode numbers, $q$, indicative of energy-dependent loss mechanisms. The pump spot size was 170~\micron. Power is estimated intracavity pump power. Error bars come from uncertainty in determining threshold and from the observed wavelength jitter during experiments.}
	\label{fig:Pcrit vs lam0}
\end{figure}

Threshold observations, as in \figref{fig:Pcrit vs lam0}, reveal that for each longitudinal mode number $q$ the critical pump power for BEC shows a minimum as a function of the cavity cutoff wavelength. Cavity cutoff is determined by the peak emission wavelength above threshold, i.e. the BEC wavelength. There are predictions that the critical pump power decreases exponentially with increasing wavelength, either with an offset~\cite{Kruchkov14} or without\cite{Kirton13}, because, close to the molecular resonance (short wavelengths), excitations are attached to dye molecules and it is only the free photons that are involved in the BEC. Going to long wavelengths, when the cavity loss rate becomes comparable to the thermalisation rate, the thermalisation breaks down and the BEC threshold pump power increases\cite{Kirton14private}. Non-radiative scattering, which occurs about once for every \mbox{$1/(1-\Phi) \simeq 20$} radiative, thermalising scattering events, becomes an important loss mechanism at short wavelengths. Since the scattering rate varies by that factor every 9~nm or so, the critical pump power varies on this scale.

At longer cavity lengths, larger $q$ values, the lowest critical pump power generally shifts to longer wavelengths. For longer cavities, the photons meet the mirrors less frequently, and hence the cavity loss rate is lower. In this way the balance between loss mechanisms shifts to longer wavelengths, where scattering events are more infrequent, although the magnitude of the shift is larger than would be expected.


{
In conclusion, we have observed dye-filled microcavity photon BEC, and seen that the macroscopic occupation of the lowest-energy state is a robust phenomenon. We have noted behaviour} which was dependent on both the pump beam geometry and the cutoff wavelength of the cavity. The critical pump power increases faster than linearly with pump spot size over the range that we have measured. The efficiency of coupling from pump light to intracavity photon number also increases with spot size, for spots smaller than the typical thermal size, implying that critical photon number increases dramatically. The size of the intracavity photon cloud also depends on pump spot size. {This evidence suggests that the pump beam couples poorly to cavity photons, but better to spatially well-matched modes, and that spatial redistribution of light is not complete.} We have also observed that critical pump power depends on ground-state-energy, with an optimum dictated by a balance between loss mechanisms: cavity photon loss and non-radiative photon scattering by the dye. We believe a model of dye-filled microcavity photon BEC that included these effects would be able to fully explain our results, and that the inclusion of these effects would render predictions of phenomena such as coherence and fluctuations more accurate.

We thank Peter Kirton, Jonathan Keeling and Jan Klaers for many informative discussions, and Lydia Zajiczek for experimental assistance. This work was funded by EPRSC fellowship EP/J017027/1.

\bibliographystyle{prsty}
\bibliography{photon_bec_refs}
\end{document}